\documentclass[12pt,a4paper]{article}

\usepackage{graphicx}% Include figure files
\usepackage{dcolumn}% Align table columns on decimal point
\usepackage{bm}% bold math

\begin{document}

\begin{center}
 
{\Large \bf Modeling multi-cellular systems using sub-cellular
elements}
 
\vspace{2cm}
 
{\large T. J. Newman}
 
\vspace{0.5cm}
 
{Department of Physics $\&$ Astronomy and School of Life Sciences, \\
Arizona State University, Tempe, AZ 85287}
                                                                               
\end{center}

\vspace{5mm}
\begin{abstract}
We introduce a model for describing the dynamics of large numbers of
interacting cells. The fundamental dynamical variables in the model
are sub-cellular elements, which interact with each other through
phenomenological intra- and inter-cellular potentials. Advantages of
the model include i) adaptive cell-shape dynamics, ii) flexible
accommodation of additional intra-cellular biology, and iii) the
absence of an underlying grid. We present here a detailed description
of the model, and use successive mean-field approximations to connect
it to more coarse-grained approaches, such as discrete cell-based
algorithms and coupled partial differential equations. We also discuss
efficient algorithms for encoding the model, and give an example of a
simulation of an epithelial sheet. Given the biological flexibility of
the model, we propose that it can be used effectively for modeling a
range of multi-cellular processes, such as tumor dynamics and
embryogenesis.
\end{abstract}

\vspace{1cm}

\newpage

\section{Introduction}
Computational modeling of multi-cellular systems has become an
increasingly useful tool for the interpretation and understanding of
experimental data in a variety of biological areas. Numerous studies
have been performed within the modeling community, with application to
collective dynamics of unicellular organisms, such as myxobacteria
\cite{igo} and slime molds \cite{mar,vas}, and to dynamics within
multicellular organisms. Studies of the latter include avascular tumor
growth \cite{byr,sto}, tumor angiogenesis \cite{cha}, embryogenesis
\cite{dav,pai}, and cell sorting \cite{dra,gran}.

A fundamental issue in modeling cell populations is that of scale. One
is typically interested in systems composed of tens of thousands to
many millions of cells, and yet the cell population is often
phenotypically heterogeneous. Therein lies the problem of whether or
not to include more biological realism at the cellular level, with the
inevitable computational cost of being limited to smaller numbers of
cells. The different modeling techniques currently employed can be
viewed as different compromises to this inescapable problem. At the
most coarse-grained level one erases cell identity and uses continuous
cell densities to describe the system. The classic model of this type
is the Keller-Segel differential equation model of aggregation in
social amoeba \cite{kel}. Similar differential equation models have
been applied to many other areas, including, of course, tumor growth
\cite{byr,cha,nag}. Cell densities can also be modeled using finite
element methods \cite{dav}. At the next finer scale, cells within the
population are modeled as discrete objects, yet with little or no
internal structure \cite{bre,new}.  Such models may be constructed
either as cellular automata on a grid, or else as many-body
simulations with no underlying lattice. Recent work has indicated that
in the presence of chemotaxis, grid effects can lead to strong
artifacts \cite{gri2}. Proceeding to smaller length scales, cells are
endowed with a size (which can change with time during the cell
cycle), and perhaps anisotropy (e.g. modeled as ellipsoids) to mimic
cell shape and/or cell polarity \cite{dra,pal}. An ingenious and
popular model of this type is that due to Graner and Glazier, in which
cells are represented as clusters of Potts spins on a fine grid
\cite{gran}. The Potts model approach has been applied to a wide range
of systems, including cell sorting \cite{gla}, slug formation in {\it
Dictyostelium} \cite{mar}, and avascular tumor growth \cite{sto}.  At
smaller scales still, the internal biochemical or biomechanical
dynamics of the cell are included, e.g. signal transduction for
response to chemical signals, and/or cytoskeleton dynamics via actin
polymerization \cite{grac,rub}.  As the scale of biological detail
becomes more fine, computational constraints limit the size of the
cell population. Indeed, models primarily concerned with cytoskeleton
dynamics typically focus on a single cell.

In this paper we introduce a framework for modeling multi-cellular
systems which is designed to allow simulation of large numbers of
cells in three dimensions, but also allows for adaptive cell shape
dynamics and the accommodation of successive degrees of intra-cellular
biology.  This framework uses ``sub-cellular elements'' (defined
below) as the fundamental dynamical variables, along with overdamped
Langevin dynamics \cite{new,van} for temporal development of the
system.

The outline of the paper is as follows. We describe the sub-cellular
element model in the next section. In section 3, we use a succession
of mean-field approximations to connect our model to more
coarse-grained descriptions. In section 4 we briefly discuss efficient
implementation of the model and give a simple example of numerical
output. We conclude in section 5 with a summary of the main results of
the paper and a discussion of biological extensions of the model.

\section{The sub-cellular element model}
The cell sets the fundamental scale in multi-cellular systems. As
such, it is natural to base model descriptions of these systems at the
cellular scale. One of the key properties to incorporate in a
cell-based model is dynamical change in cell shape (or more generally,
cell polarity), which can occur in response to local mechanical
interactions with neighboring cells, or in response to long-ranged
chemical signaling \cite{alb}. Adaptive changes in cell shape/polarity
allow coherent dynamics of large numbers of cells. For example,
several mechanisms of large-scale morphological change during
gastrulation are due to cell intercalation, which is driven by
elongation of individual cells along a particular axis \cite{wol}.
From a modeling perspective, 
cell shape is difficult to parameterize. For instance, systematically
extending an ellipsoid model of cells in three dimensions requires
complicated geometrical constructions. Ideally, one would like a model
in which cell
shape emerges from cellular interactions -- in other words, for cell
shape to be adaptive to the local environment. Here, we attempt to
instantiate this property by sub-dividing each cell into a number of
``sub-cellular elements.'' Both the intra- and inter-cellular dynamics
are written in terms of interactions between these elements. We shall
first describe this dynamics by writing the equations of motion for
the elements, and then discuss how this dynamics can be interfaced
with the underlying biology.

For simplicity, consider a system with a constant number $N$ of
cells in three spatial dimensions, with each cell being
composed of $M$ elements. We label an individual cell by $i \in (1,N)$
and an element in cell $i$ by $\alpha _{i} \in (1,M)$.  For ease of
discussion we assume that chemical signaling is absent from the system,
so that cells respond purely to local biomechanical interactions. In
this case, the position vector of element $\alpha _{i}$ is taken to
change in time according to three processes: i) a weak stochastic
component, which mimics the underlying fluctuations in the dynamics of
the cellular cytoskeleton, ii) an elastic response to intra-cellular
biomechanical forces, and iii) an elastic response to inter-cellular
biomechanical forces. We assume further that the elements' motion is
over-damped, so that inertial effects can be ignored. The equation of
motion for the position vector of element $\alpha _{i}$ takes the
form:
\begin{equation}
\label{elementeq}
{\dot {\bf y}}_{\alpha _{i}} = {\bf \eta}_{\alpha _{i}}
-\nabla _{\alpha _{i}}\sum \limits _{\beta _{i} \ne \alpha _{i}}
V_{\rm intra}(|{\bf y}_{\alpha _{i}}-{\bf y}_{\beta _{i}}|)
-\nabla _{\alpha _{i}}
\sum \limits _{j \ne i}\sum \limits _{\beta _{j}}
V_{\rm inter}(|{\bf y}_{\alpha _{i}}-{\bf y}_{\beta _{j}}|) \ .
\end{equation}
On the right-hand-side, the noise term ${\bf \eta}_{\alpha _{i}}$ is
a Gaussian distributed random variate with zero mean and correlator
\begin{equation}
\label{noise}
\langle {\bf \eta}^{m}_{\alpha _{i}}(t){\bf \eta}^{n}_{\beta _{j}}(t') 
\rangle = 2\nu \delta _{i,j}\delta _{\alpha _{i},\beta _{j}} \delta ^{mn}
\delta (t-t') \ ,
\end{equation}
where $m$ and $n$ are vector component labels in the three-dimensional
space.  The second and third terms on the right-hand-side of
Eq. (\ref{elementeq}) represent, respectively, intra- and
inter-cellular interactions between the elements. These interactions
are completely characterized by the phenomenological potentials
$V_{\rm intra}$ and $V_{\rm inter}$.  At this level of description,
all relevant biological detail must be encoded into these two
potentials. The elemental composition of cells, along with the
inter-elemental potentials, are shown schematically in Fig. 1.
We have assumed that ``two-body'' potentials are sufficient to 
describe the dynamics. It may be necessary to use ``three-body''
potentials to capture the essence of more complicated interactions.

For given biological applications of this modeling framework, one must
intuit (or better, derive) reasonable forms for $V_{\rm
intra}$ and $V_{\rm inter}$. For illustrative purposes, consider a
population of cells which are weakly adhesive to one
another. Sub-cellular elements both within and between cells will be
mutually repulsive if their separation is below the equilibrium size
of an element. For separations larger than this size, the elements
will be mutually attractive, but with the strength of attraction falling off
rapidly with separation. These properties can, for example, be
conveniently encoded via a generalized form of the Morse potential,
which is commonly used in physics and chemistry to model
inter-molecular interactions \cite{shi}.  The (generalized) Morse
potential has the explicit form
\begin{equation}
\label{morse}
V(r)=U_{0}\exp (-r/\xi_{1}) - V_{0}\exp (-r/\xi_{2}) \ ,
\end{equation}
and is illustrated in Fig. 2. It is straightforward to evaluate the
position and depth of the attractive potential minimum in terms of the
four parameters $(U_{0},V_{0},\xi _{1},\xi _{2})$. In a simple
application of the element model, one can use Morse potentials for
both $V_{\rm intra}$ and $V_{\rm inter}$, with parameters chosen to
ensure that the former has stronger inter-elemental adhesion than the
latter. This condition is necessary, in this simplest version of the
model, to maintain the mechanical integrity of the cells.

The introduction and explicit choice of these potentials has so far
been purely phenomenological, and some discussion of the biological
motivation for these potentials is necessary. Considering a
``typical'' tissue cell, such as a fibroblast or epithelial cell, the
mechanical integrity of the cell is maintained by the internal
cytoskeleton \cite{alb}. This is a complex network of different types
of interconnected filaments, with actin being the most important
filament type for cell motility.  Dividing the cell into elements
corresponds to modeling the shape and mechanical integrity of the cell
in terms of volume elements of cytoskeleton. The intra-cellular
attraction between elements arises from the mechanical rigidity of the
cytoskeleton, more specifically, the elastic forces transmitted
through filaments connecting neighboring elements. These interactions
are local and thus it is necessary that the potential has a rapid
decay with distance. Elements at opposite sides of a cell {\it
mechanically} interact through elastic forces mediated by elements
comprising the interior of the cell. The biochemical and biomechanical
interactions between cells is complex, and arises from a variety of
cell-cell (and cell-matrix) contacts, such as gap, tight, and anchoring
junctions \cite{alb}. Still, the interactions are local, and, once the
cells are linked, one can think of the interaction in terms
of a short-ranged elastic potential. There is no reason to favor the
Morse potential at this level of description -- there are many
reasonable potentials that one can write down. Such potentials will,
however, be characterized by at least four parameters -- two energy
scales (for short-ranged repulsion, and intermediate-range adhesion)
and two length scales, which characterize the size of an element, and
its adhesive range.  There are a number of models which have been
focused on the detailed mechanics of the cytoskeleton, and its role in
cell motility \cite{dim,grac,mog,rub}. An interesting subject of
future study is the derivation of inter-elemental potentials from
coarse-graining the underlying cytoskeletal mechanics considered in
these more detailed studies.

\section{Connections to coarse-grained models}

In this technical section we sketch the derivation of coarser-grained
models by applying a succession of mean-field approximations to the
element model. The essence of this section is summarized in Fig. 3,
which shows the fundamental objects/fields characterizing the cell
models at different scales.

In the first of these coarse-graining steps, we replace the element
model by a sub-cellular density model, in which the discrete elements
within a given cell $i$ are replaced by a smooth average density field
$\rho _{i}({\bf x},t)$. We stress that a separate density field exists
for each cell in the system, although the density fields are strongly
correlated to one another. To proceed, we first recast the
sub-cellular element model in terms of the probability distribution of
individual elements. We define the probability distribution of element
$\alpha _{i}$ by $P_{\alpha _{i}}({\bf x},t)=\langle \delta ^{3}({\bf
x}-{\bf y}_{\alpha _{i}}) \rangle $, where the angled brackets denote
an average over the noise $\eta $. Starting from
Eqs. (\ref{elementeq}) and (\ref{noise}) we use standard methods
\cite{new,van} to derive an equation of motion for $P_{\alpha _{i}}$,
which takes the form
\begin{eqnarray}
\label{palphai}
\nonumber
\partial_{t} P_{\alpha _{i}}({\bf x},t)= & & \nu \nabla ^{2}
P_{\alpha _{i}}({\bf x},t)\\
\nonumber
& + & 
\nabla \cdot \int d^{3}x' \ \left [ \nabla V_{\rm intra}(|{\bf x}-{\bf x}'|)
\right ] \sum \limits _{\beta _{i} \ne \alpha _{i}}
P_{\alpha _{i},\beta _{i}}({\bf x},t;{\bf x}',t) \\
& + & 
\nabla \cdot \int d^{3}x' \ \left [ \nabla V_{\rm inter}(|{\bf x}-{\bf x}'|)
\right ] \sum _{j \ne i} \sum \limits _{\beta _{j}}
P_{\alpha _{i},\beta _{j}}({\bf x},t;{\bf x}',t) \ ,
\end{eqnarray}
where $P_{\alpha _{i},\beta _{j}}$ is the ``two-element'' distribution
function.  The equation of motion for this two-element distribution
will involve the three-element distribution, and so on. The simplest
truncation scheme to break the hierarchy of equations is the mean
field approximation (MFA), in which the statistical correlations
between elements are discarded. Within this MFA we have $P_{\alpha
_{i},\beta _{j}}({\bf x},t;{\bf x}',t)= P_{\alpha _{i}}({\bf
x},t)P_{\beta _{j}}({\bf x}',t)$.

We now define the sub-cellular density of cell $i$ via $\rho _{i}({\bf
x},t) =\sum _{\alpha _{i}}P_{\alpha _{i}}({\bf x},t)$. Summing over
$\alpha _{i}$ in Eq. (\ref{palphai}), and imposing the MFA, we find a
closed equation for this sub-cellular density function, which takes
the form of an advection-diffusion equation:
\begin{equation}
\label{subcellade}
\partial _{t}\rho _{i}({\bf x},t) = \nu \nabla ^{2}\rho _{i}({\bf x},t)
+\nabla \cdot \rho _{i}({\bf x},t)\nabla \Phi _{i}({\bf x},t) \ ,
\end{equation}
where the velocity potential experienced by the density field of 
cell $i$ is given by
\begin{equation}
\label{velpoti}
\Phi _{i}({\bf x},t)=\int d^{3}x' \ V_{\rm intra}(|{\bf x}-{\bf x}'|)
\rho _{i}({\bf x}',t)
+ \int d^{3}x' \ V_{\rm inter}(|{\bf x}-{\bf x}'|)
\sum _{j \ne i}\rho _{j}({\bf x}',t) \ .
\end{equation}
The MFA used to derive this density equation will typically be good
when the number of elements used to define the cell is very large.
The density representation may well be interesting to explore from an
analytical standpoint. However, it is probably not so useful for
numerical implementation. For simulation of $N$ cells, one must
simultaneously integrate $N$ coupled partial differential equations on
a fine three dimensional grid. As we shall see in the next section,
the underlying element model, expressed in Eq. (\ref{elementeq}), can
be very efficiently encoded for simulation with no need of an
underlying grid. 

We take this opportunity to mention that Eq. (\ref{subcellade}) can be
discretized in such a way that it resembles a master equation
\cite{gri1}. In this representation, the density of cell $i$ can be
interpreted as a probability distribution of identical elements, which
move from one grid site to the next by activated hopping. The hopping
rate has the Arrhenius form $\sim \exp (-\Delta \Phi _{i}/2\nu)$
(where $\Delta \Phi _{i}$ is the change in velocity potential, for 
an element from cell
$i$, between the two grid sites of interest, which is actually
non-trivial to compute since it depends self-consistently on the
density $\rho _{i}$).  The multi-cellular system as a whole is then
defined on a grid, with each grid site able to accommodate (one or
more) elements from the $N$ different cells. The elements move about
the lattice (and consequently interact) via activated hopping, with
highly non-linear hopping rates as indicated above. This
representation, although not easily implemented, illustrates a
qualitative connection between a discretized form of the sub-cellular
element model (after one level of MFA) and the lattice-based Potts
model \cite{gran}.  This discretized form of the element model has the
flavor of a lattice-gas analog to the Potts model -- in the sense that
a lattice-based element moves over the lattice and yet keeps its
original parent cell identity, whereas a Potts spin is defined at a
lattice site and identifies, at a given time, the cell spanning that
particular site. Having to hand these two distinct models of
multi-cellular systems (the sub-cellular element model, as expressed in
Eq.(\ref{elementeq}), and the Potts model) defined at similar scales of
biological realism, will allow useful cross-validation of these
approaches, especially when applied to complicated biological systems.

We can use the density equation (\ref{subcellade}) to coarse-grain to
another scale -- where now only gross properties (which we refer to
loosely as ``moments'') of the sub-cellular density field are used to
characterize the cell. This coarse-graining step is analogous to a
multipole expansion in electromagnetism. The zeroth moment of cell $i$
is its mass, which is defined by $m_{i}(t)=\int d^{3}x \ \rho
_{i}({\bf x},t)$. Within the present discussion, this quantity is
independent of time and cell index $i$ since we have assumed that all
cells have the same number of elements, and that the number of
elements does not change with time. Proceeding to the first moment, we
define the position vector of the center of mass of cell $i$ via ${\bf
x}_{i}(t) = \int d^{3}x \ {\bf x} \rho _{i}({\bf x},t)$. The equation
of motion for this position vector is obtained from
Eq. (\ref{subcellade}) and takes the form
\begin{equation}
\label{com}
{\dot {\bf x}}_{i} = -\int d^{3}x \ \rho _{i}({\bf x},t)
\nabla \Phi _{i}({\bf x},t) \ .
\end{equation}
We briefly mention the second moment of cell $i$, namely its inertia
tensor \cite{lan}. This is defined via $T_{i}^{mn}(t) = \int d^{3}x \
\left ( x^{2}\delta ^{mn} - x^{m}x^{n} \right ) \rho _{i}({\bf x},t)$.
An equation of motion, similar to Eq. (\ref{com}), can be written down
for this tensor. This quantity contains crucial information regarding
the {\it mechanical polarity} of the cell. Higher order moments can
be defined and contain successively more information about the shape
and density distribution of the cell.

Since the equations of motion of these moments are written in terms
of integrals over the sub-cellular density, they are all strongly
inter-dependent. To write closed equations again requires some form of
truncation. Here we use the simplest, which is again a form of MFA,
namely: $\int d^{3}x \ f({\bf x})\rho _{i}({\bf x},t) = f({\bf
x}_{i}(t))$. This allows us to express the right-hand-side of
Eq. (\ref{com}) in terms of ${\bf x}_{i}(t)$, and we find the closed
equation
\begin{equation}
\label{discrete}
{\dot {\bf x}}_{i}(t) = -\nabla \sum _{i \ne j}
V_{\rm inter}(|{\bf x}_{i}(t)-{\bf x}_{j}(t)|) \ .
\end{equation}
There are two interesting points to note: i) the intra-cellular
potential has vanished under this approximation, since we are
essentially shrinking the cells to points, and ii) the dynamics are
now deterministic.  Concerning the first point, the intra-cellular
potential will reappear in this coarse-grained description if we
include second-order effects -- namely, if we derive two coupled
equations for each cell, describing the time-dependence of the cell's
position vector and its inertia tensor.  Concerning the second point,
the effect of the noise $\eta _{\alpha_{i}}$ has vanished since the
first MFA leading to Eq. (\ref{subcellade}) essentially assumes an
infinite number of elements, so that the explicit noise terms are
averaged to zero. The noise from a finite number ($M$) of elements
will be non-zero, and have a variance which scales as $1/M$.  This
weak noise (which describes the random wandering of the center of
mass) can be added to Eq. (\ref{discrete}) {\it a posteriori} in order
to retain stochasticity in this discrete cell representation.  One can
then write Eq. (\ref{discrete}) as
\begin{equation}
\label{discrete_noise}
{\dot {\bf x}}_{i}(t) = {\bf \eta}_{i}(t)-\nabla \sum _{i \ne j}
V_{\rm inter}(|{\bf x}_{i}(t)-{\bf x}_{j}(t)|) \ ,
\end{equation}
where the noise $\eta _{i}$ has zero mean, and correlator $\langle
\eta ^{m}_{i}(t)\eta ^{n}_{j}(t') \rangle = 2D \delta _{i,j}\delta
^{mn}\delta (t-t')$, where $D=\nu /M$.  This stochastic model, which
tracks the positions of the cells, is precisely that studied by Newman
and Grima \cite{new}. As shown in that work, a further MFA applied to
Eq. (\ref{discrete_noise}) leads to a closed equation for the density
of cells, which is defined via $n({\bf x},t) = \sum _{i} \langle
\delta ^{3} ({\bf x}-{\bf x}_{i}(t)) \rangle $. We omit the details
here and simply give the final result:
\begin{equation}
\label{deneq}
\partial _{t} n({\bf x},t) = D \nabla ^{2} n({\bf x},t)
+\nabla \cdot n({\bf x},t)\nabla \Psi ({\bf x},t) \ ,
\end{equation}
where the coarse-grained velocity potential $\Psi $ for the cell density
has the form 
\begin{equation}
\label{velpot}
\Psi ({\bf x},t)=\int d^{3}x' \ V_{\rm inter}(|{\bf x}-{\bf x}'|)
n({\bf x}',t) \ .
\end{equation}
A rigorous derivation of this last step has been given by Stevens
in the context of chemotaxis \cite{ste}, and uses the limit of
infinite cell number, with an appropriately scaled chemotactic
coupling.
 
Finally, after three levels of coarse-graining, we have arrived at a
partial differential equation for the cell density, as given in
Eqs. (\ref{deneq}) and (\ref{velpot}). As mentioned in the
Introduction, this level of description has been widely used to
describe the large-scale dynamics of cell populations. However, as
should be clear from this analysis, a great deal of statistical
information and smaller-scale biomechanics must be discarded at this
scale. It would be very interesting to rederive the density equations
from a more careful analysis. Some details of the intra-cellular
potentials (especially regarding cell polarity) can be captured in
this largest-scale description through i) calculating renormalized
parameters, such as the diffusion coefficient $D$, in the density
description (\ref{deneq}), and ii) deriving the companion equation for
a ``cell polarity field'' from the discrete cell equations for the
inertia tensor.

\section{Efficient algorithms and model output}

We now return to the sub-cellular element model as described in
section 2.  The numerical implementation of this model turns out to be
fairly straightforward and efficient. Since the fundamental dynamical
variables are position vectors, we have no need for an underlying
grid, and simply need to track the values of the $(M\times N)$ vectors
$\lbrace {\bf y}_{\alpha _{i}} \rbrace $ which completely describe the
state of the system at any given time.

It is worth mentioning that some care must be taken in constructing
the algorithm so as to avoid a CPU cost which scales as $(MN)^{2}$.
This would arise from attempting to interact every element with every
other in order to update the system. Clearly this is not
necessary since the potentials are short-ranged. As such, the
algorithm only needs to interact a given element $\alpha _{i}$ with
those elements which are close enough to have a non-negligible
interaction. So long as we can efficiently identify these nearby
elements, our algorithm will have a CPU cost which scales as $MN$.
This will allow the simulation of large numbers of cells with moderate
to large numbers of elements per cell. There
are a number of ways to identify nearby elements. The methods to
achieve this have been developed over the years in molecular dynamics
simulations \cite{hai,rap}.  Examples are neighbor tables and the more
sophisticated binary search trees and octrees. We have employed a
method based on ``sectors.''  The three dimensional system is broken
up into a grid of sectors, the size of a sector chosen to be about
twice the range of inter-elemental interactions. The dynamics of the
elements are completely oblivious to the sectors.  The sectors simply
allow one to construct a look-up table wherein for each sector there
is a list of the identity of those elements in that sector. When one
calculates the interactions of a given element, one computes the
interactions between that element and all those in its own sector and
the nearest-neighbor sectors. Temporal development is performed using
an explicit Euler discretization, with time increment $\delta t$.

In Fig. 4 we show an example of the output of this algorithm. In this
example we have simulated 128 cells, with each cell constructed from
20 sub-cellular elements. Both the intra- and inter-cellular
potentials are chosen to be generalized Morse potentials, with parameter sets
$(U_{0},V_{0},\xi _{1},\xi _{2})=(0.25,0.1,0.12,0.36)$ and
$(0.25,0.05,0.12,0.24)$ respectively.  Other parameter values are $\nu
= 0.001$ and $\delta t = 0.1$.  For ease of presentation we have
simulated the cells in a quasi-two-dimensional geometry, with
hard-wall boundary conditions at $z=0$ and $z=0.5$, which could
represent an epithelial layer bounded by a basement membrane. The system
is shown after about 3000 iterations, starting from a random
distribution of cell positions (with the initial positions of the
elements for each cell randomly distributed in a small region about
these cell positions).  This simulation requires about 3 minutes on a 2GHz
PC. Extrapolating to larger systems, we see that thousands of
iterations for a system of $10,000$ cells (each with 20 elements)
requires a few hours of CPU time on a PC. More sophisticated
optimization of the algorithm will allow further improvements in
efficiency.  Note that as a result of the elements attempting to form
inter-cellular adhesive bonds with elements of nearby cells, the cells
have adapted their cell shapes to their local biomechanical
environment. A more systematic numerical study comparing the different
coarse-grained descriptions is currently in progress \cite{smi}.

\section{Summary and outlook}

In this paper we have introduced a model of interacting multi-cellular
systems, in which the fundamental objects are not cells, but
``sub-cellular elements'' -- by which we mean, in the simplest sense,
small volume elements of intra-cellular cytoskeleton. The dynamics of
the elements is described by Langevin equations, as given in
Eq. (\ref{elementeq}). The three dynamical contributions to a given
element's motion are i) a weak stochastic component, ii) local
biomechanical interactions with other elements within the same cell,
described by a phenomenological potential $V_{\rm intra}$, and iii)
local biomechanical interactions
with elements in nearby cells, these described by a potential $V_{\rm
inter}$ (see Fig. 1). The success of the model in a given biological
application depends, in large part, on well-chosen forms of these
potentials. The generic form of these potentials is reasonably well
captured by the generalized Morse potential (Fig. 2).  We have outlined, in
section 3, a series of coarse-graining procedures (summarized in
Fig. 3), whereby the sub-cellular element model can be linked to
models at larger scales, such as discrete cell models and differential
equation models describing the macroscopic cell density. In section 4
we indicated how an efficient algorithm may be constructed to
integrate Eq. (\ref{elementeq}) forward in time for large numbers of
cells/elements, and gave an example from a simulation of a sheet of cells
(Fig. 4).

We end the paper with a discussion of some of the many possible
extensions of the model, whereby biological detail can be added
without distortion of the underlying element framework.

\subsection{Cell types}
For simplicity, in this introductory paper we have presented the
element model for a population of $N$ identical cells.  Phenotypic
heterogeneity at the cell level can be described by attaching cell
labels to the noise strength ($\nu \rightarrow \nu _{i}$),
intra-cellular potentials ($V_{\rm intra} \rightarrow V_{i}$), and
inter-cellular potentials ($V_{\rm inter} \rightarrow V_{i,j}$).

\subsection{Sub-cellular element types}
For many applications it will not be sufficient to construct a cell
from $M$ identical elements. For instance, it might be necessary to
distinguish elements in the interior of the cell (``cytoplasmic
elements'') from elements on the surface of the cell (``membrane
elements''). Different element types can easily be instantiated by
defining the appropriate classes of intra- and inter-cellular
potentials. In the example just given, the inter-cellular interactions
would primarily be described by an inter-cellular potential connecting
membrane elements from neighboring cells, while the cytoplasmic
elements would interact only with elements within the same cell.  In
addition, elements can be endowed with internal variables registering
environmental variables such as pH or nutrient level, and communicate
these variables to neighboring elements to trigger appropriate cell
response.

\subsection{Extra-cellular element types}
Cells not only adhere to each other, but also interact biomechanically
with a range of non-cellular environmental structures, such as the
extra-cellular matrix or various gel-like media \cite{alb}.  In some
cases these structures are produced by the cells themselves. Within
the spirit of the element model, these extra-cellular structures can
also be constructed from elements (i.e. elastically coupled degrees of
freedom on the same length scale as the sub-cellular elements) with
appropriately defined interaction potentials to describe the
cell/non-cell interactions.

\subsection{Chemotaxis}
We have been exclusively concerned with short-ranged biomechanical
interactions in this paper. Equally important for many applications
(e.g. embryogenesis, wound healing) are long-ranged chemical
interactions. The simplest way to introduce such interactions in the
element model is to chemically couple the centers of mass of cells
which are signaling to one another. Then, since the source and sink of
the signal are points, it is straightforward to implement the Green
function methods developed in some detail in Newman and Grima
\cite{new}. These methods encode the diffusion of chemical signals
through the diffusion equation Green function (which allows one to
avoid introducing a fine grid for explicit integration of the chemical
diffusion equations). More sophisticated treatments would be based on
cell polarity induced by the chemical signal. For instance, one could
use chemically responsive element types within the cell, so that the
cell as a whole responds to a chemical signal via a sub-cellular
element response, followed by a cell-level response mediated by
elastic interactions of the responsive element to the non-responsive
ones.

\subsection{Cell cycle}
For many applications, and in particular that of tumor growth, cells
in the population are undergoing numerous cell divisions over the time
scales of interest. Cell growth can be accommodated in the modeling
framework by allowing new elements to be spawned within the cell (based
on, for instance, an internal monitoring of nutrient levels as
discussed above in subsection 5.2). Mitosis is more complicated.  It
can be ``forced'' upon a cell using some form of threshold conditions,
under which the elements segregate into two daughter cells.  However,
more elegant mechanisms can no doubt be devised.

\vspace{0.3cm}

In this final section we have tried to give a flavor of possible
biological extensions of the model, and the ease with which they can
be implemented within the element framework. This flexibility must
ideally be tempered by minimal incremental model-building in order to
keep models simple enough to understand from both quantitative and
biological perspectives. With this caveat in mind, we hope that the
sub-cellular element model will prove useful in the computational
study of a wide range of multi-cellular systems.

\vspace{0.5cm}

The author would like to thank James Glazier, Ramon Grima, John Nagy,
Kevin Schmidt, Erick Smith, and Cornelius Weijer for interesting
discussions. The author gratefully acknowledges partial support from
the NSF (IOB-0540680) and NSF/NIH (DMS/NIGMS-0342388).

\newpage

\begin{figure}
\includegraphics[width=\linewidth]{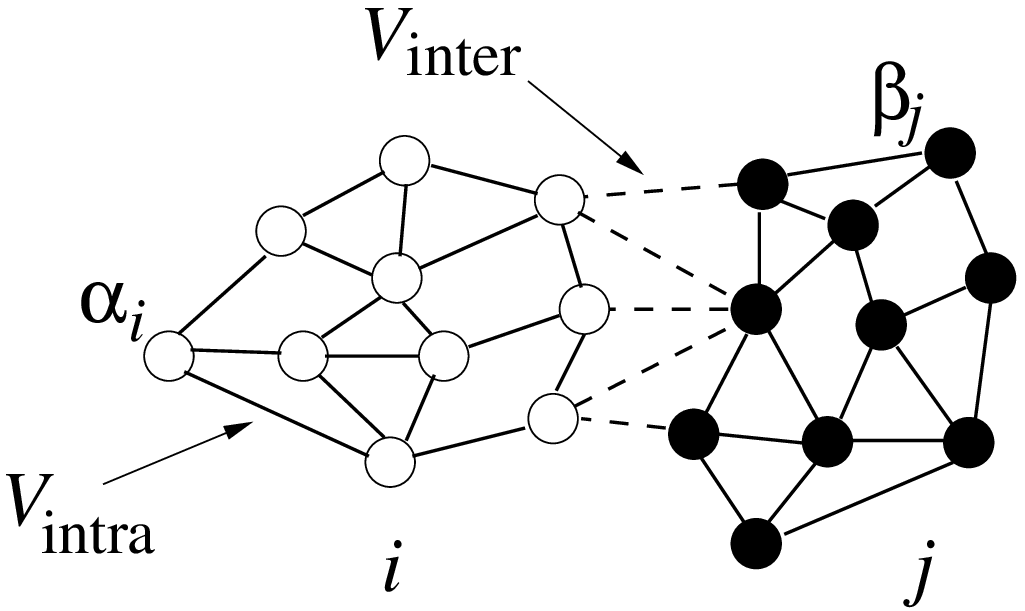}
\caption{Schematic diagram showing two cells, $i$ and $j$, and a subset of the
intra- and inter-cellular interactions between their elements. The
elements of cell $i$ are represented by open circles, and those of
cell $j$ by filled circles. The intra- and inter-cellular interactions
are represented by solid and dashed lines respectively.}
\end{figure}

\vspace{2.0in}

\begin{figure}
\includegraphics[width=\linewidth]{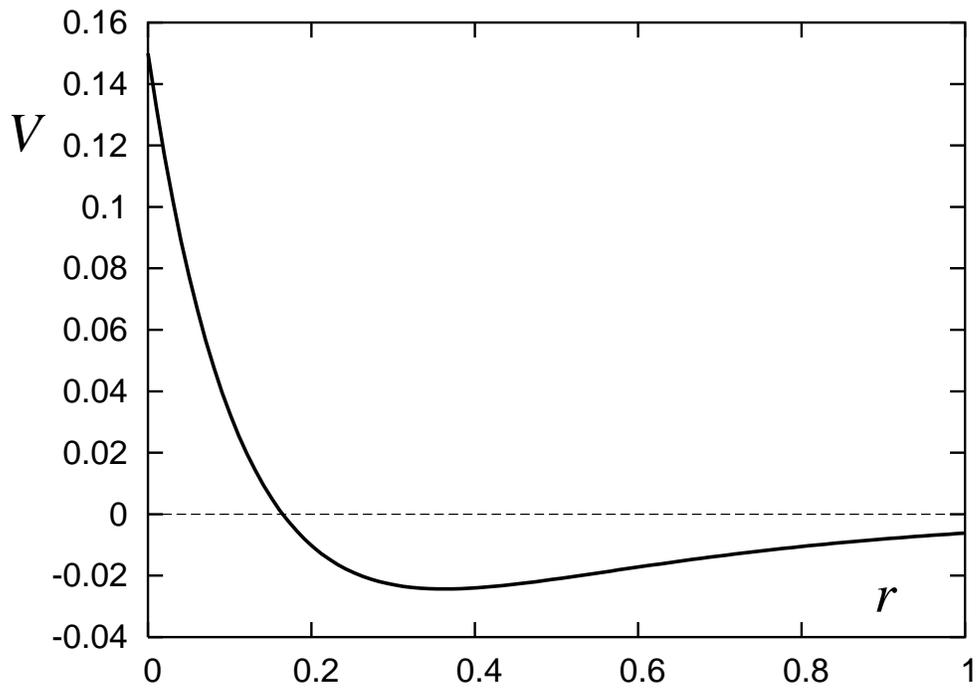}
\caption{The Morse potential for parameter values $U_{0}=0.25$, $V_{0}=0.1$,
$\xi _{1}=0.12$, and $\xi _{2}=0.36$, as used for the intra-cellular
potential in section 4.}
\end{figure}

\begin{figure}
\includegraphics[width=\linewidth]{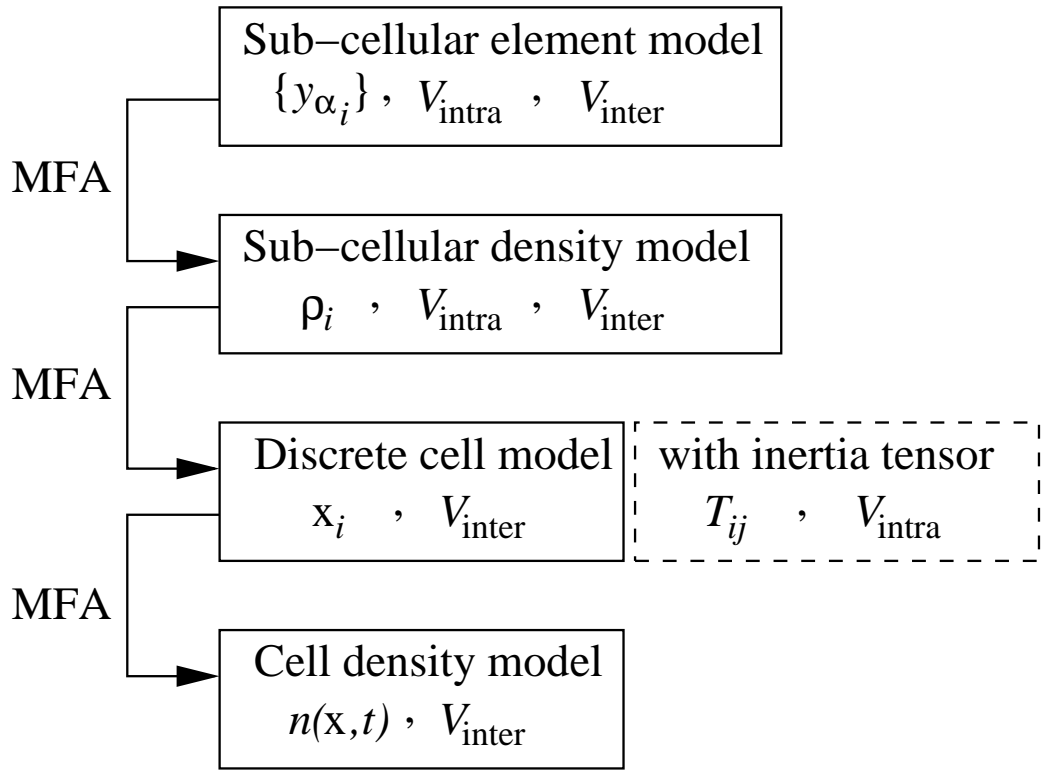}
\caption{Relationships between model descriptions of multi-cellular
systems at different scales. The four levels (from finer to coarser
scales) are described explicitly by Eqs. (\ref{elementeq}),
(\ref{subcellade}), (\ref{discrete_noise}), and (\ref{deneq})
respectively.}
\end{figure}

\newpage

\begin{figure}
\includegraphics[width=\linewidth]{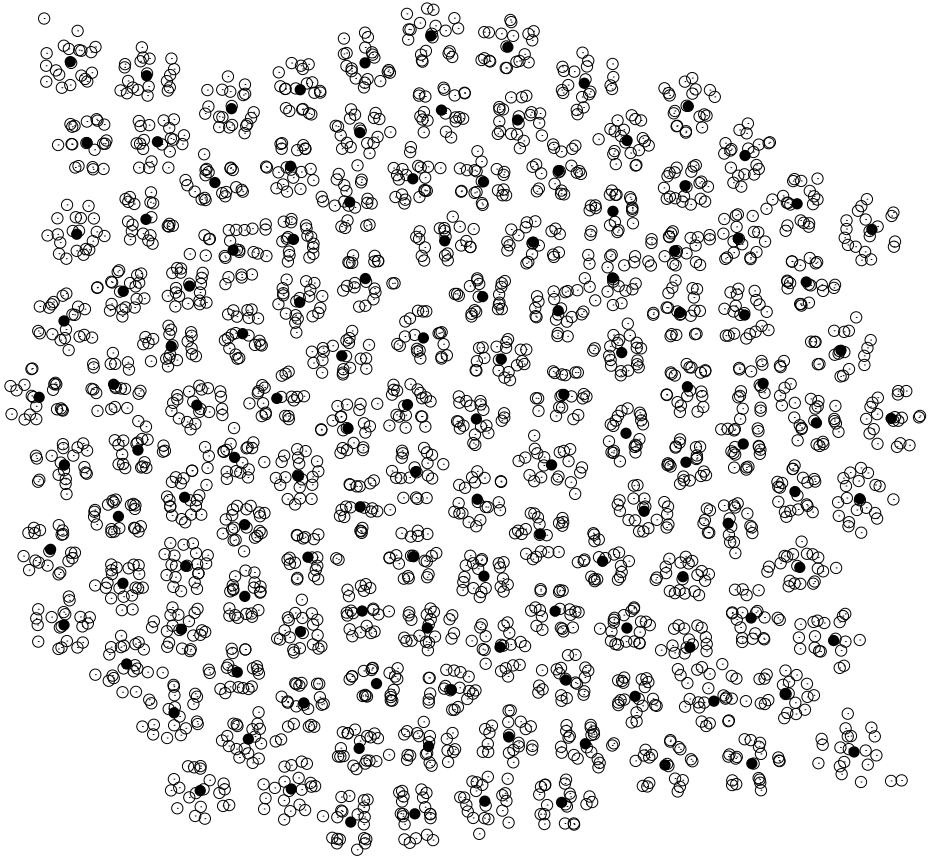}
\caption{An example of numerical output from the model using Morse
potentials. Parameter values are given in the main text. Shown here is
a two-dimensional {\it projection} of data from a model of an epithelial
sheet.  Each of the 128 cells is composed of 20 elements (open
circles), and element motion is constrained in the third dimension by
hard-wall boundaries. The filled circles indicate the center of mass
of each cell.}
\end{figure}

\end{document}